\documentclass[9pt,twocolumn,twoside]{pnas-new}
\usepackage[final]{pdfpages}
\usepackage{verbatim}

\templatetype{pnasmathematics} 

\title{Range contraction enables harvesting to extinction}

\author[a,b,c,d,1]{Matthew G. Burgess}
\author[a,b,c]{Christopher Costello} 
\author[b]{Alexa Fredston-Hermann}
\author[e]{Malin L. Pinsky}
\author[a,b,c]{Steven D. Gaines}
\author[b,d,1]{David Tilman}
\author[d,f]{Stephen Polasky}

\affil[a]{Sustainable Fisheries Group}
\affil[b]{Bren School of Environmental Science and Management}
\affil[c]{Marine Science Institute, University of California, Santa Barbara, CA 93106, USA}
\affil[d]{Department of Ecology, Evolution and Behavior, University of Minnesota, St. Paul, MN 55108, USA}
\affil[e]{Department of Ecology, Evolution and Natural Resources, Rutgers University, New Brunswick, NJ 08901}
\affil[f]{Department of Applied Economics, University of Minnesota, St. Paul, MN 55108, USA}

\leadauthor{Burgess} 

\significancestatement{Many threatened species including elephants, sturgeons, and bluefin tunas are harvested for high-value products. Species can be driven extinct if incentives to harvest do not diminish as populations decline; this occurs if harvest prices rise faster than costs with declining stock. Whereas recent conservation attention for these species has largely focused on market demand, we show--using a theoretical model and an empirical review--that contractions in species’ geographic ranges, which stabilize costs and may be especially common among terrestrial species, might often play a larger role in maintaining harvest incentives. Forces impacting ranges--such as patchy and declining habitats, schooling/herding behavior, and climate change--therefore merit greater attention in assessing overharvesting threats.}

\authorcontributions{M.G.B. designed research; M.G.B., C.C., A.F.-H.,M.L.P.,S.D.G.,D.T., and S.P. performed research and wrote the paper; M.G.B., A.F.-H., and M.L.P analyzed the data.}
\authordeclaration{The authors declare no conflicts of interest.}
\correspondingauthor{\textsuperscript{1}To whom correspondence should be addressed. E-mail: mburgess@ucsb.edu or tilman@umn.edu}

\keywords{Anthropogenic Allee Effect $|$ hyperstable $|$ endangered species $|$ poaching} 

\begin{abstract}
Economic incentives to harvest a species usually diminish as its abundance declines, because harvest costs increase. This prevents harvesting to extinction. A known exception can occur if consumer demand causes a declining species’ harvest price to rise faster than costs. This threat may affect rare and valuable species, such as large land mammals, sturgeons, and bluefin tunas. We analyze a similar but underappreciated threat, which arises when the geographic area (range) occupied by a species contracts as its abundance declines. Range contractions maintain the local densities of declining populations, which facilitates harvesting to extinction by preventing abundance declines from causing harvest costs to rise. Factors causing such range contractions include schooling, herding, or flocking behaviors--which, ironically, can be predator-avoidance adaptations; patchy environments; habitat loss; and climate change. We use a simple model to identify combinations of range contractions and price increases capable of causing extinction from profitable overharvesting, and we compare these to an empirical review. We find that some aquatic species that school or forage in patchy environments experience sufficiently severe range contractions as they decline to allow profitable harvesting to extinction even with little or no price increase; and some high-value declining aquatic species experience severe price increases. For terrestrial species, the data needed to evaluate our theory are scarce, but available evidence suggests that extinction-enabling range contractions may be common among declining mammals and birds. Thus, factors causing range contraction as abundance declines may pose unexpectedly large extinction risks to harvested species.
\end{abstract}

\dates{Accepted draft: see published version at www.pnas.org/cgi/doi/10.1073/pnas.1607551114}
\doi{\url{www.pnas.org/cgi/doi/10.1073/pnas.1607551114}}

\begin{document}

\verticaladjustment{-2pt}

\maketitle
\thispagestyle{firststyle}
\ifthenelse{\boolean{shortarticle}}{\ifthenelse{\boolean{singlecolumn}}{\abscontentformatted}{\abscontent}}{}

\dropcap{H}arvesting has driven the population declines of thousands of species of animals and plants \citep{international2015iucn}, but it is thought to rarely cause extinction because the increasing cost of harvesting a progressively rarer species would eventually exceed the value of the harvest, and harvesting would stop \citep{clark1976mathematical}. However, for species harvested for high-value products, there is concern that their depletion could fuel price increases, via market demand, large enough to compensate for higher harvest costs and thereby maintain profit-incentives to harvest all the way to extinction, absent management intervention \citep{courchamp2006rarity}. Courchamp \textit{et al}. \citep{courchamp2006rarity} term this threat the `Anthropogenic Allee Effect'. 

Species thought to face this threat include those harvested, both legally and illegally, for trophies (e.g., large terrestrial mammals including rhinoceros, elephants and large cats \citep{graham2011biodiversity,palazy2011cat,palazy2012rarity, biggs2013legal,wittemyer2014illegal}), for collections (e.g., stag beetles \citep{tournant2012rarity}), for body parts regarded as having medicinal or aphrodisiac properties (e.g., many large mammals \citep{graham2011biodiversity}), or for luxury foods (e.g., sturgeons, bluefin tunas, sea cucumbers \citep{gault2008consumers,collette2011high,purcell2014cost}). Many such species are considered threatened by the Red List of the International Union for Conservation of Nature (IUCN) \citep{international2015iucn} or the Convention on International Trade in Endangered Species of Wild Fauna and Flora (CITES) \citep{cites2016}. Expanding human populations, coupled with economic growth in developing countries with large luxury harvest markets, may increase pressures on these species in coming decades \citep{purcell2014cost}. 

A similar threat of extinction from overharvesting would occur if a species' harvest costs failed to rise as its abundance declined, thus maintaining harvesting profitability. One way this can occur is if the geographic area (range) occupied by the species contracts as its abundance declines, thereby maintaining its local population density. This pattern has been noted in several fish and aquatic invertebrate populations \citep{prince2008contraction,wilberg2009incorporating} and may have contributed to the famous 1990s collapse of northern cod (\textit{Gadus morhua}) \citep{rose1999hyperaggregation}. In fish and invertebrates, range contraction is often observed in declining populations that exhibit schooling behavior (to maintain school sizes) and/or forage in patchy environments (because populations concentrate in the preferred habitats) \citep{prince2008contraction,wilberg2009incorporating,thorson2016density}. Habitat destruction and climate change can also cause range contraction, and thus might similarly buffer harvest costs against population declines and create incentives for harvesting to extinction.

These overharvesting threats from range contraction and market demand likely interact (Fig. \ref{fig:1}). For example, prices would not need to be very sensitive (`flexible') to abundance declines to allow profitable harvesting to extinction if costs were insensitive to abundance declines because of range contraction. 

Here, we theoretically and empirically characterize this interaction. We use a simple model to illustrate conditions under which range contraction and price flexibility can in combination allow harvesting to extinction under open access. We then review available empirical evidence in order to shed light on where these biological and economic risk factors may be most acute. While many harvests are now managed \citep{harris2013application}, it is nonetheless important to understand the threats posed by open-access incentives due to the pervasiveness of illegal and unreported harvesting of endangered species \citep{graham2011biodiversity,biggs2013legal,wittemyer2014illegal}. 

\section*{Theory}
We consider the following model of open-access harvesting on a single population with abundance $N$. The population has a per-capita growth rate, denoted $g(N)$, which follows negative density-dependence ($g'[N] < 0$) and has a maximum of $r$ ($g[0] = r$). The population is harvested at rate $Y$, which is a function of the current harvest effort, $E$, and abundance ($N$); there is no harvest if either abundance or effort is zero:
\begin{equation}
Y=Y(N,E),Y(0,E)=Y(N,0)=0\mbox{.}
\label{eq:1}
\end{equation}
The population's rate of change, denoted $\dot{N}\equiv dN/dt$, is: 
\begin{equation}
\dot{N}=Ng(N)-Y\mbox{.}
\label{eq:2}
\end{equation}
Under open access, the rate of change in effort, $\dot{E}\equiv dE/dt$, has the same sign as harvest profits \citep{clark1976mathematical}, i.e.:
\begin{equation}
\dot{E}\mbox{ >(<) }0\mbox{ if } p\mbox{ >(<) }c \mbox{, }\dot{E}=0\mbox{ if } p=c\mbox{,}
\label{eq:3}
\end{equation}
where $p$ and $c$ are respectively the price and average cost of a unit of harvest received/incurred by harvesters. 

\subsection*{Extinction condition}
Harvesting to extinction can occur in this model if and only if, as abundance ($N$) approaches 0, there is still upward pressure on effort (i.e. $\dot{E}>0$) when harvest rates ($Y$) exactly balance population growth ($Ng[N]$) (i.e. abundance is not changing, $\dot{N}=0$) (illustrated in \textit{Supporting Information} [SI] Fig. S1). Formally, this means that extinction can occur if and only if:
\begin{equation}
\lim_{N \to 0} \left(\frac{p}{c}\right)|_{Y=Ng(N)}>1\mbox{.}
\label{eq:4}
\end{equation}

\begin{figure}
\centering
\includegraphics[width=.4\linewidth]{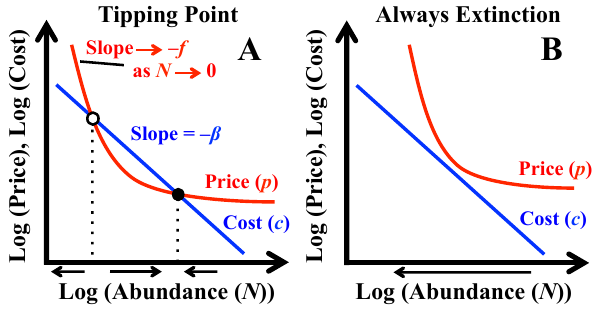}
\caption{Profitable harvesting to extinction (Courchamp \textit{et al.}'s \citep{courchamp2006rarity} `Anthropogenic Allee Effect') under open access occurs when, at harvest levels at which abundance is not changing, harvest price (red, nonlinear due to density-dependent population growth) is greater than harvest cost (blue) as abundance approaches zero. Such conditions either can result in alternative stable states (panel \textbf{A})--a tipping-point abundance (open circle) separates domains of attraction of a positive equilibrium abundance (filled circle) and extinction--or can cause profits to be positive at any abundance and make extinction the only possible outcome of open-access harvesting (panel \textbf{B}).}
\label{fig:1}
\end{figure}

\subsection*{Prices, costs, and abundance}
We assume that the sensitivity of the price ($p$) to changes in abundance ($N$) is mediated by changes in the supply of harvest in the market--equal to the harvest rate, $Y$. The sensitivity of a harvest product's price to its supply is commonly measured by the 'price flexibility of demand' \citep{houck1965relationship,eales1997generalized}, denoted $f$, defined as:
\begin{equation}
\frac{\partial{p}}{\partial{Y}}\left(\frac{Y}{p}\right)=-f\mbox{,}
\label{eq:5}
\end{equation}
i.e. price increases by $f$ percent when supply ($Y$) declines by one percent. Price flexibility is related to, but distinct from, the more widely known concept of demand elasticity; they are reciprocals in a market with only one good, but not otherwise \citep{houck1965relationship}. We use price flexibility, because it is considered a more appropriate empirical demand measure for harvests (e.g. \citep{eales1997generalized}). 

Assuming constant price flexibility ($f$), eq. (\ref{eq:5}) implies that the price ($p$) is given by:
\begin{equation}
p=\rho Y^{-f}\mbox{,}
\label{eq:6}
\end{equation}
where $\rho$ is either a constant or a function of variables other than supply ($Y$). In the analysis that follows, we assume $\rho$ is constant, but discuss alternate assumptions below in \textit{Other considerations} and in the \textit{SI Materials and Methods}.

To model the relationship between average costs ($c$) and abundance ($N$), we assume the harvest rate ($Y$) at a given time is proportional to the harvesting effort ($E$) (i.e. constant returns to scale) multiplied by the population abundance ($N$) raised to a constant power, $\beta$ (\textit{sensu} \citep{harley2001catch,clark1975economics}): 
\begin{equation}
Y=qN^{\beta}E\mbox{;}
\label{eq:7}
\end{equation}
$q$ is also a constant. $\beta$ represents the percent change in catch-per-unit effort (CPUE $=Y/E$) resulting from a one-percent change in abundance ($N$), and thus can be thought of as the 'catch flexibility'. If $\beta < 1$, then CPUE is 'hyperstable' because it changes proportionally more slowly than abundance \citep{harley2001catch}.

For simplicity we assume that effort has a constant cost, and that the units of effort are such that they have unit costs (i.e. total cost = $E$). The average unit cost of harvest ($c = E/Y$), is then given by [from eq. \ref{eq:7}]:
\begin{equation}
c=q^{-1}N^{-\beta}\mbox{.}
\label{eq:8}
\end{equation}
We briefly discuss alternate assumptions in \textit{Other considerations} and in the \textit{SI Materials and Methods}.

\subsection*{Extinction condition revisited}
With equations \ref{eq:6} and \ref{eq:8} for $p$ and $c$, extinction condition \ref{eq:4} becomes:
\begin{equation}
\lim_{N \to 0} \left(\frac{p}{c}\right)|_{Y=Ng(N)}= \left\{
\begin{array}{ll}
      0 & f < \beta \\
      \rho qr^{-f} & f = \beta \\
      \infty & f > \beta \\
\end{array} 
\right\rbrace > 1\mbox{,}
\label{eq:9}
\end{equation}
implying (Fig. S1; intuition discussed below and illustrated in Fig. \ref{fig:1}) that extinction can occur if and only if:
\begin{equation}
f > \beta \mbox{, or } f = \beta \mbox{ and } \rho qr^{-f}>1\mbox{.}
\label{eq:10}
\end{equation}
If price flexibility exceeds catch flexibility ($f > \beta$), then two scenarios are possible (depending on parameter values) under open access: (i) extinction is the only possible outcome (Fig. \ref{fig:1}B; light-red $E$ isocline in Fig. S1D); (ii) there is an unstable equilibrium acting as a tipping point separating a basin of attraction of extinction and a basin of attraction of either a stable (positive) equilibrium or limit cycle (Fig. \ref{fig:1}A; dark-red $E$ isocline in Fig. S1D). In contrast, if $f = \beta$ and $\rho qr^{-f}>1$, only extinction is possible (light-red $E$ isocline in Fig. S1E). 

\subsection*{Range contraction}
Under random search, CPUE would be proportional to population density (abundance [$N$]/range area [$A$]), i.e. $Y \propto \left(\frac{N}{A}\right)E$, and catch flexibility ($\beta$) would then be determined solely by the relationship between abundance ($N$) and range area ($A$): $\beta =1-\frac{\partial{A}}{\partial{N}}\left(\frac{N}{A}\right)$. In practice, other factors (e.g., technology, harvester search skill) tend to further buffer CPUE as abundance declines (\citep{pitcher1995impact,wilberg2009incorporating}; see Fig. \ref{fig:2}, Dataset S1 for many examples), meaning that,
\begin{equation}
\beta \leq 1-\frac{\partial{A}}{\partial{N}}\left(\frac{N}{A}\right)\mbox{.}
\label{eq:11}
\end{equation}

\subsection*{Intuition}
The intuition of our theory is as follows: A one-percent decrease in abundance ($N$) of a rare species increases average costs by $\beta$ percent and increases price by approximately $f$ percent. If $f > \beta$, the incentive to keep harvesting only increases; if $f = \beta$, it fails to decrease (and it is positive if $\rho qr^{-f}>1$). If range ($A$) contracts proportionally as fast as or faster than abundance declines (i.e. $\frac{\partial{A}}{\partial{N}}\left(\frac{N}{A}\right) \geq 1 \Rightarrow \beta \leq 0$), then incentives to harvest are maintained even with constant prices ($f = 0$), and range contraction therefore poses an extinction threat on its own \citep{pitcher1995impact,rose1999hyperaggregation}. Similarly, if prices rise proportionally as fast or faster than supply falls (i.e. $f \geq 1$), then incentives to harvest are maintained even with no range contraction (i.e. $\beta = 1$), and price flexibility poses an extinction threat on its own. The most common case, as we will see below, is where $0 < \beta \mbox{, } f < 1$--one where price flexibility and range contraction (and/or other factors making costs insensitive to declines) can only pose threats in combination.

\subsection*{Other considerations}
Other factors beyond price and catch flexibilities can impact the changes in costs or prices coincident with changes in abundance. These include economies of scale (meaning  $\frac{\partial{Y}}{\partial{E}}\left(\frac{E}{Y}\right)>1$) (e.g. \citep{torres2014productivity}), technology and technological change (meaning $dq/dt > 0$) \citep{squires2013technical}, supply-independent increasing trends in price (e.g., caused by income growth) (meaning $\partial{\rho}/\partial{t} > 0$), and supply-independent effects of abundance on prices caused by 'rarity value' \citep{courchamp2006rarity,hall2008endangering}) ($\partial{\rho}/\partial{N}<0$). 

These other effects are not the focus of the present study, but we show in \textit{SI Materials and Methods} the direction in which each of these effects is likely to impact extinction threats: Technological change, positive supply-independent price trends, and rarity value each exacerbate the threats (though supply-independent rarity effects on prices may be negligibly small in comparison to supply effects [measured by $f$] for species already rare enough for their harvest products to confer status; e.g. see Fig. S2). Economies of scale have a neutral effect when $f \approx 1$ or $\beta \approx 1$ but can have a mild mitigating effect when $f$ and $\beta$ are both small (Fig. S3). We briefly discuss this effect in context with observed economies of scale in fisheries (Table S1). We also show how the existence of perfect substitutes (i.e. goods that are indistinguishable from the focal species' harvest product to consumers) will tend to prevent profit-driven extinction altogether. Finally, we discuss the implications of non-constant $f$ and $\beta$--in short, extinction still requires $f \geq \beta$ (assuming constant $\rho$ and $q$, and $\frac{\partial{Y}}{\partial{E}}\left(\frac{E}{Y}\right)=1$), at the limits of $f$ and $\beta$ as abundance ($N$) approaches zero. 

Other factors--such as intertemporal discounting \citep{clark1973profit,swanson1994economics} and opportunistic or multispecies harvesting \citep{branch2013opportunistic,burgess2013predicting}--can influence a species' risk of extinction by overharvesting via different mechanisms, but these are also not the focus of this study.

\section*{Review of Empirical Evidence}
Our theory suggests that profitable harvesting to extinction requires highly flexible prices, highly inflexible catch rates (implying inflexible costs), or a combination ($f \geq \beta$, which may be a conservative criterion, given the exacerbating factors not included in our model). We assume that price flexibility does not directly influence--nor is influenced by--either catch flexibility or the range-abundance relationship, and thus we review empirical estimates of these three factors separately (Fig. \ref{fig:2}), rather than restricting our analysis to species having estimates of each (there are very few such species).  

\subsection*{Price flexibility}
To our knowledge, price flexibility ($f$) estimates for wild terrestrial harvests are rare, perhaps because few terrestrial harvests of wild animal species are legal and commercial in data-rich countries. One modeling study \citep{hertzler2004african}, however, assumed $f  = 0.1$ for poached elephants. A second study \citep{rentsch2013prices} estimated demand elasticity for Serengeti bushmeat to be > 1, which is consistent with $f < 1$ with few substitutes. 

\begin{figure}
\centering
\includegraphics[width=.6\linewidth]{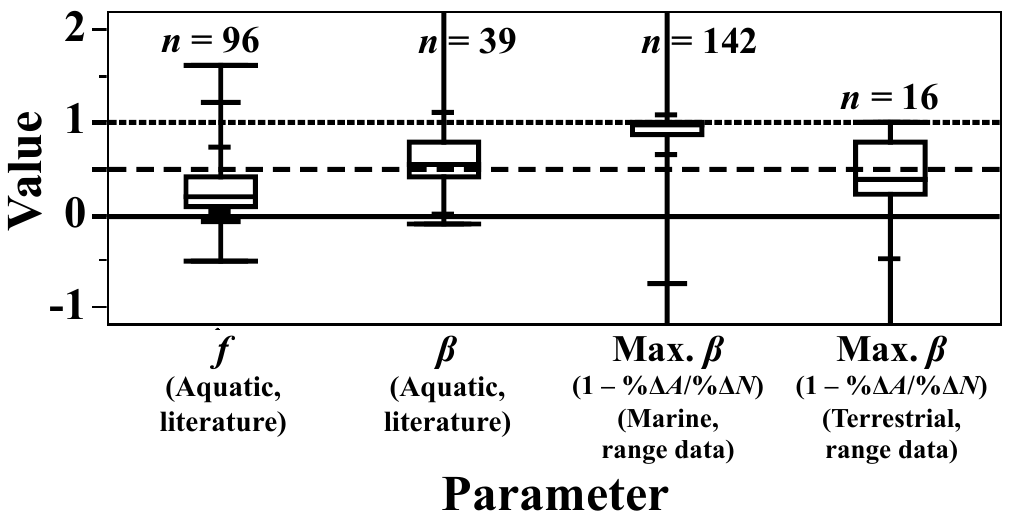}
\caption{Comparison of ranges of estimates of price flexibility ($f$: the percent increase in price, $p$, caused by a one-percent decrease in harvest rate, $Y$) and catch flexibility ($\beta$: the percent decrease in CPUE caused by a one-percent decrease in abundance, $N$) from the published literature for aquatic species; and observed range-abundance relationships (which imply upper bounds on catch flexibility [Max. $\beta$] by inequality \ref{eq:11}) in marine fish and invertebrates, terrestrial mammals and one terrestrial bird species (northern bobwhite). Boxes show 25th-75th percentile range; minima, maxima, and 2.5th,10th, 90th, and 97.5th percentiles are marked on the whiskers. See Dataset S1 for all values and references. Two terrestrial mammal populations having no observed abundance change are excluded.}
\label{fig:2}
\end{figure}

In contrast, we found 96 published price-flexibility estimates for aquatic harvest products (Fig. \ref{fig:2}, Dataset S1)--varying widely in focal species, commodity type, position in the supply chain (ex-vessel [the price paid to harvesters], wholesale, retail) and estimation method, including some from high-value species. Most estimated price flexibilities were low (median $f = 0.22$; $f < 0.5$ in 81\% of estimates). Estimates of ex-vessel price flexibilities--of primary relevance to our theory--were especially small ($n = 46$, median = 0.12, $f < 0.5$ in 96\% of estimates) (Dataset S1). Ex-vessel prices tend to be less flexible than prices further along the supply chain \citep{asche2007studies}, perhaps because of market power among some wholesale buyers and processors, or because storage and preservation can buffer retail supplies against changes in harvest rates. 

\begin{figure}
\centering
\includegraphics[width=0.5\linewidth]{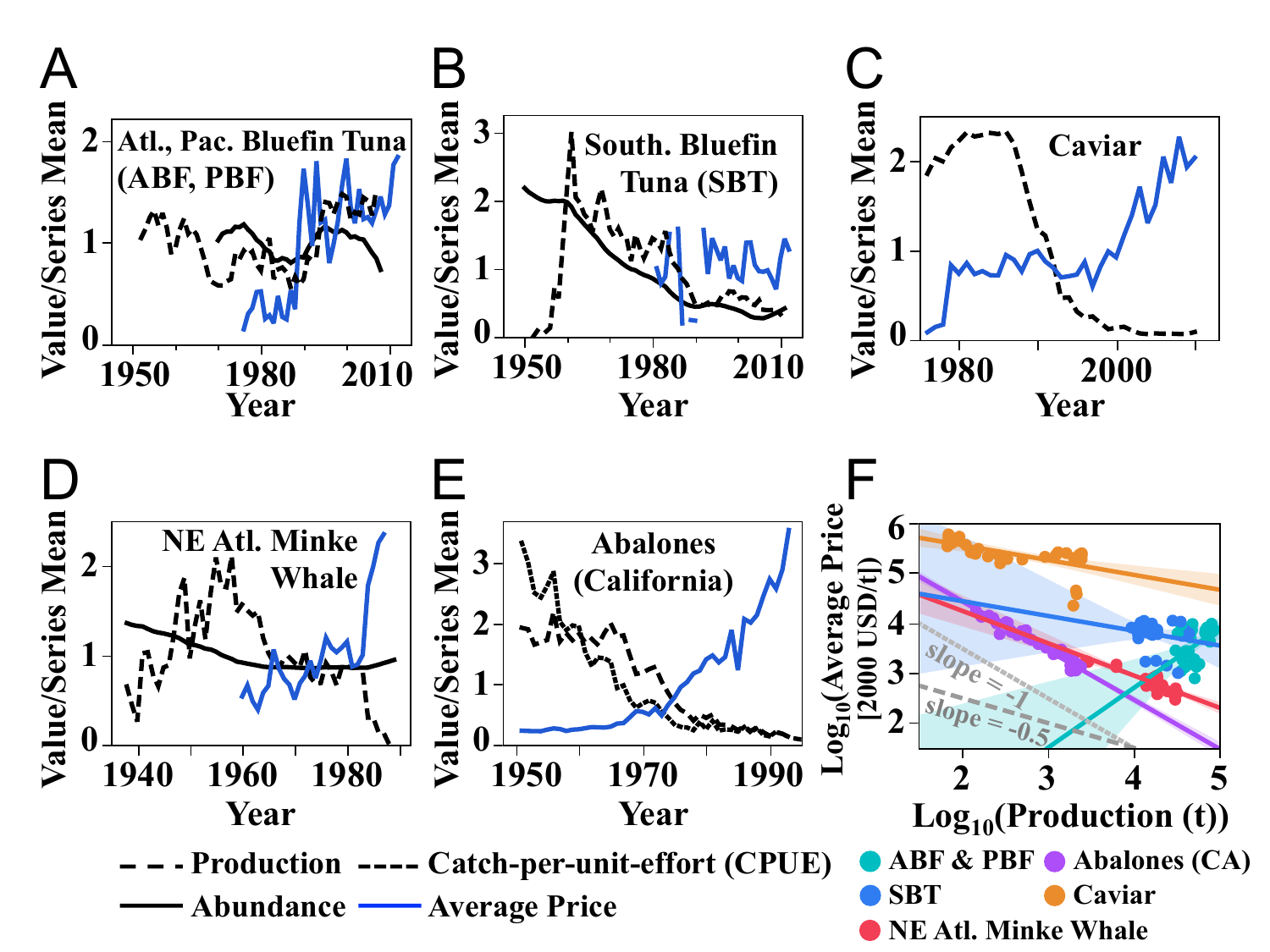}
\caption{Price trends in highly valued marine species: (\textbf{A}) ABF and PBF (production [total catch] and abundance from ref. \citep{ricard2012examining}; ex-vessel prices from ref. \citep{melnychuk2016reconstruction}), (\textbf{B}) SBT (production [total catch] and abundance from ref. \citep{ricard2012examining}; ex-vessel prices from ref. \citep{melnychuk2016reconstruction}), (\textbf{C}) caviar (global production, average export prices from ref. \citep{faofish2016}), (\textbf{D}) the Northeast Atlantic minke whale (production, abundance, prices from ref. \citep{amundsen1995open}), and (\textbf{E}) California abalones (CPUE and prices from refs. \citep{courchamp2006rarity,hobday2000status}). All prices were converted to real USD value using the World Bank's \citep{wdi2016} published currency exchange and inflation rates. (\textbf{F}) Prices of each of these harvests have historically risen as fast as catch has declined (CA abalone) or slower (others). Solid lines show linear fits of log-transformed price and production (supply) data, with 95\% confidence intervals shaded. Dotted and dashed grey lines respectively illustrate slopes of -1 (implying 1\% increase in price for 1\% decrease in production) and -0.5, for reference.}
\label{fig:3}
\end{figure}

Some high-value aquatic species lack formal price flexibility estimates from in-depth demand analyses, but their price trends are nonetheless mostly consistent with inflexible prices ($f << 1$). To illustrate this point, Fig. \ref{fig:3} shows aggregate price-supply relationships for several populations thought to be facing (or recently have faced) demand-related extinction threats. Bluefin tunas (Atlantic [ABF] [\textit{Thunnus thynnus}], Pacific [PBF] [\textit{Thunnus orientalis}], and Southern [SBT] [\textit{Thunnus maccoyii}]) have had relatively stable prices despite abundance declines (Fig. \ref{fig:3}A,B; data from refs. \citep{melnychuk2016reconstruction,ricard2012examining}). Indeed, Chiang et al. \citep{chiang2001impact} estimated $f = 0.19$ for wholesale fresh bluefin tuna (lumped ABF, PBF, and SBT) in Japan. Historical caviar (made from sturgeon [Acipenseriformes] roe) prices (1976-2010; Fig. \ref{fig:3}C; data from ref. \citep{faofish2016}) have risen ~0.3\% on average for every 1\% decline in catch. Before the International Whaling Commission’s moratorium on whaling in the late 1980s, Northeast Atlantic minke whale (\textit{Balaenoptera acutorostrata}, a relatively data-rich whale example; data from ref. \citep{amundsen1995open}) prices increased by ~0.65\% on average for every 1\% decline in catch [Fig. \ref{fig:3}D]). California abalone (\textit{Haliotis sp.}) prices seem to have been more flexible, increasing by roughly 1\% (0.98\%) for every 1\% decline in catch over the period 1950-1993, before the 1997 ban on fishing (Fig. \ref{fig:3}E) (data from refs. \citep{courchamp2006rarity,hobday2000status}). These historical trends do not necessarily reflect the magnitudes of underlying market-level price flexibilities, but they are mostly consistent with the pattern of low price flexibility ($f < 1$) seen in other aquatic harvest products (Fig. \ref{fig:3}F). Controlling for catch, none of these species have residual prices significantly correlated with abundance (Fig. S2), suggesting that supply-independent rarity effects may indeed be negligible.

\subsection*{Range contraction and catch flexibility}
Several studies have directly estimated catch flexibility ($\beta$; usually incorporating effects of both range and other factors) in aquatic populations (see ref. \citep{wilberg2009incorporating} for review), but to our knowledge no such literature yet exists for terrestrial populations. We found published estimates of catch flexibility from 39 aquatic populations (median $\beta = 0.56$) (Fig. \ref{fig:2}; see Dataset S1 for values and references). Of these, 32 (82\%) exhibited hyperstability ($\beta < 1$), and 14 (36\%) exhibited severe hyperstability ($\beta < 0.5$), including 5 of 6 small pelagic finfish populations--which tend to both exhibit schooling behavior and forage in patchy environments. 

\begin{figure*}
\centering
\includegraphics[width=0.6\textwidth]{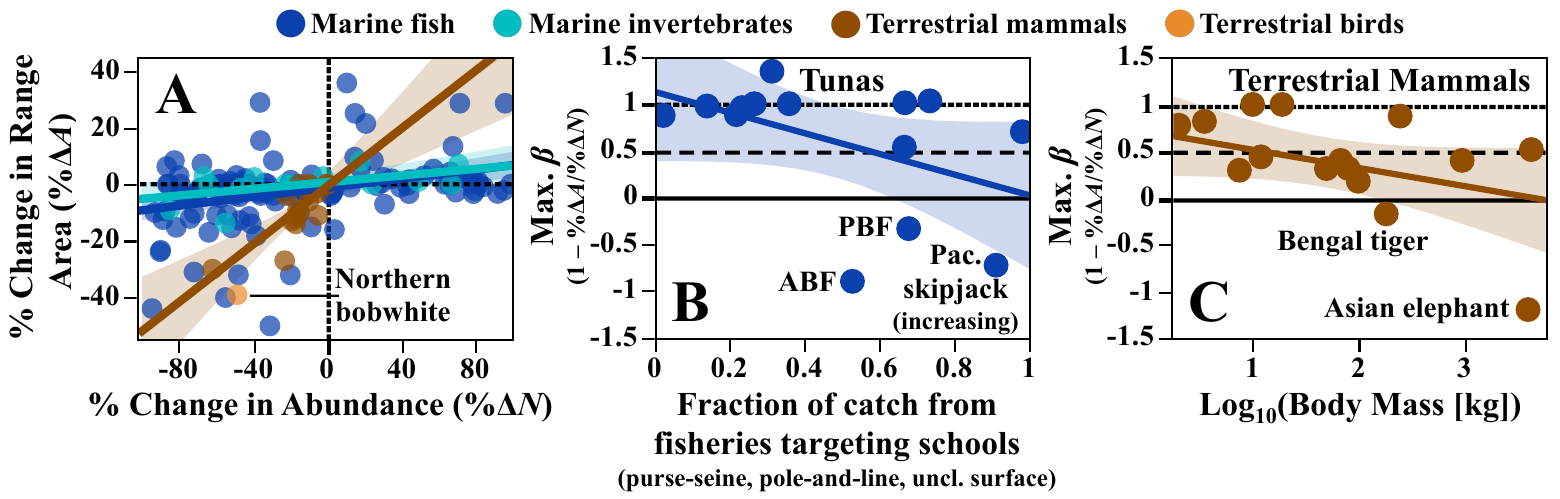}
\caption{Range-abundance relationships for populations of: harvested U.S. marine fish and invertebrates (1970s-2000s; from ref. \citep{pinsky2013marine}), tuna and billfish (1960-1999; from refs. \citep{ricard2012examining,worm2011range}), harvested terrestrial mammals, mostly from the Western Ghats of India (1978/79-2008/09; from ref. \citep{pillay2011patterns}; also the African forest elephant (\textit{Loxodonta cyclotis}) from 2002-2011, ref. \citep{maisels2013devastating}), and northern bobwhite (\textit{Colinus virginianus}), a North American game bird (1966-1993; from ref. \citep{rodriguez2002range}). Panel \textbf{A} shows all populations. Panels \textbf{B} and \textbf{C} show relationships between the maximum catch flexibility (Max. $\beta$) and: (\textbf{B}) a proxy for schooling behavior in tunas, and (\textbf{C}) adult body mass in terrestrial mammals (see Dataset S1). Harvested populations exhibiting hyperaggregation (Max. $\beta$ < 0) are labeled; all are declining in abundance and range over the time periods in question, except for Pacific skipjack tuna (\textit{Katsuwonus pelamis}), which is increasing. Colored lines in all panels represent linear ordinary least squares (OLS) fits (within large taxonomic groups in panel \textbf{A}), with 95\% confidence intervals shaded. Negative slopes in panels \textbf{B} and \textbf{C} are nearly significant ($p<0.1$).}
\label{fig:4}
\end{figure*}

For comparison, we compiled estimates of coincident changes in range area and abundance in 142 harvested marine fish and invertebrate populations (Fig. \ref{fig:4}A; see caption for references; full data available in Dataset S1). We use the ratio of coincident observed percent changes in range and abundance, denoted $\%\Delta A/\%\Delta N$, as a measure of average $\frac{\partial{A}}{\partial{N}}\left(\frac{N}{A}\right)$, which would bound catch flexibility ($\beta$) by inequality \ref{eq:11}. 

Most of these populations had very abundance-insensitive ranges (median $\%\Delta A/\%\Delta N=0.03$) (Fig. \ref{fig:2},\ref{fig:4}A): for 87 (61\%) of them, the range changed by less than one-tenth of the percent by which abundance changed ($\left|\%\Delta A/\%\Delta N\right|<0.1$) (see also ref. \citep{thorson2016density}). We found $\%\Delta A/\%\Delta N>0.5$ (implying $\beta < 0.5$) in only 8 (6\%) of the populations, half of which were tunas or billfish (Dataset S1), and we found some evidence suggesting that abundance-sensitive range may be associated with schooling behavior among tunas (Fig. \ref{fig:4}B). Notably, we found hyperaggregation ($\%\Delta A/\%\Delta N>1$, implying $\beta < 0$) in both ABF and PBF (Fig. \ref{fig:4}B), suggesting that these populations would not necessarily need any price flexibility to be profitably harvested to extinction. 

In the \textit{SI Materials and Methods}, we compare the range-abundance relationships from these 142 harvested marine populations to a taxonomically similar sample from 247 non-harvested marine populations. In the combined sample, we find taxonomy, but not harvesting, to be a significant determinant of the range-abundance relationship (\textit{SI Materials and Methods}, Tables S2, S3, Figs. S4, S5), suggesting that ecology may be a more important driver. We most commonly find highly abundance-sensitive ranges ($\%\Delta A/\%\Delta N>0.5$) among small invertebrates (molluscs, shrimps) and pelagic finfish in the combined sample (Table S3).

Our reviews of catch flexibility estimates and range-abundance relationships (Dataset S1) support the hypothesis that aggregation and patchy habitats lead to hyperstable catches in aquatic species, in part because of range contraction \citep{prince2008contraction,wilberg2009incorporating}. However, we find catch flexibility ($\beta$) estimates to be smaller on average than values implied by range-abundance relationships alone (Fig. \ref{fig:2}), which suggests that other drivers besides range contraction also contribute to hyperstable catch rates. Indeed, a few studies have directly demonstrated this (e.g., refs. \citep{rose1999hyperaggregation,erisman2011illusion} found $\frac{\partial{\mbox{CPUE}}}{\partial{D}}\left(\frac{D}{\mbox{CPUE}}\right)\approx 0.5$, where $D=N/A$, in Atlantic cod and California kelp bass [\textit{Paralabrax clathratus}] respectively; see refs. \citep{pitcher1995impact,wilberg2009incorporating} for review). 

We found estimates of coincident range and abundance trends for 17 harvested terrestrial mammal and one North American game bird population (northern bobwhite [\textit{Colinus virginianus}]) (Fig. \ref{fig:4}A; see caption for description of sources; full data available in Dataset S1). In most (10) of these populations, range contracted at more than half the rate abundance declined ($\%\Delta A/\%\Delta N>0.5$, implying $\beta < 0.5$) (Fig. \ref{fig:4}A,C). Abundance-sensitive range was most common among large-bodied mammals; we found hyperaggregation in the Bengal tiger (\textit{Panthera tigris}) and the Asian elephant (\textit{Elephas maximus})--both species threatened by poaching and habitat destruction \citep{international2015iucn,pillay2011patterns} (Fig. \ref{fig:4}C). 

Though this terrestrial sample is quite small, it suggests that terrestrial species might more commonly experience range contraction as they decline than marine species--a pattern which we also find in a sample of 28 non-harvested bird and mammal populations (Fig. S5, \textit{SI Materials and Methods}, Dataset S1). If, like for aquatic species, other factors besides range contraction further decrease catch flexibility ($\beta$) (Fig. \ref{fig:2}), many terrestrial species--especially large-bodied mammals, perhaps--could be susceptible to profitable harvesting to extinction under open access.

\section*{Discussion}
We find that escalating prices, stable harvest costs, or combinations of these two factors, can, in theory, allow a species to be profitably harvested to extinction, absent effective protection. Our empirical review suggests that stable harvesting costs might be a surprisingly common cause of such extinction risks, but this full range of possibilities merits much further attention. Our review suggests that: (i) range contraction in declining species may be common, and often severe enough to allow extinction with little or no price flexibility, among terrestrial species, and among aquatic species that school or forage in patchy habitats; (ii) high price flexibility also occurs among some highly valued aquatic species; and (iii) other factors besides range contraction also buffer harvest costs against abundance declines. These results suggest that risk factors for abundance-insensitive costs merit greater attention in harvested species conservation. 

Low price flexibility may be common because most harvested commodities have partial substitutes in the market. For example, studies have found moderate-to-high market substitutability between species within broad classes of fish products (see ref. \citep{asche2007studies} for review), including between most tuna species \citep{chiang2001impact,sakai2009substitute}. Species whose harvests have perfect substitutes are unlikely to be profitably harvested to extinction \citep{hall2008endangering}, because perfect substitutes cause the abundance-sensitivity of a species' price to diminish as it approaches extinction (see \textit{SI Materials and Methods}). However, our price analyses were restricted to commercially harvested aquatic species, so the pattern of inflexibility we observed may not necessarily extend to other types of markets (e.g. black markets).    

Our review suggested that range changes may be more correlated with abundance changes among terrestrial species compared to marine species (Figs. \ref{fig:4}A, S4, S5). This difference may be related to greater habitat destruction and space limitation on land. Space limitation positively correlates with body size in terrestrial mammals \citep{mcnab1963bioenergetics,ripple2014status}, which may explain the especially strong relationships between range and abundance changes we found in large-bodied mammals. Thus, habitat destruction is likely a key driver of abundance-insensitive costs on land; and it also may directly and independently add to the extinction risk faced by terrestrial species. However, given our relatively small and unrepresentative terrestrial samples, these hypotheses merit further scrutiny.  

Aggregating in patchy habitats is common among schooling fish, many small aquatic invertebrates, and other taxa found to have low catch flexibility, as well as some birds. Aggregation and habitat patchiness are likely key drivers of these species’ susceptibility to both range contraction as they decline (due to concentration in preferred habitats and/or need to maintain school size \citep{prince2008contraction,wilberg2009incorporating}) and to some other socioeconomic factors making costs even less sensitive to abundance (e.g. aggregation predictability \citep{pitcher1995impact,wilberg2009incorporating}, harvester coordination [e.g. \citep{gaertner2004analysis}], use of fish aggregating devices [FADs] \citep{torres2014changes}). It is ironic that herding and schooling behaviors, which are considered adaptations for predator avoidance \citep{hamilton1971geometry}, may make such species particularly prone to extinction from human harvesting. 

Together, these results suggest that the harvested species most susceptible to profitable harvesting to extinction may be those with aggregation behavior, patchy or declining habitats, large home ranges (on land), and few market substitutes \citep{hall2008endangering}. Moreover, our results suggest that--absent management--many high-value harvested species could potentially be threatened by these risk factors. These include Atlantic and Pacific bluefin tuna (for which we find evidence of hyperaggregation [i.e. $\beta < 0$]), and many large poached land mammals (which likely have low catch flexibility); and may include sturgeons (which face habitat destruction, e.g. \citep{lenhardt2006assessment}) and high-value marine invertebrates (e.g., abalones, for which we found evidence suggestive of high price flexibility [Fig. \ref{fig:3}E,F]). The extinct passenger pigeon (\textit{Ectopistes migratorius}), which aggregated in large flocks and suffered both habitat destruction and overharvesting (e.g. \citep{hung2014drastic}), possibly provides a historical case study of the interaction of these risk factors in extinction. The susceptibility of some plants (e.g. orchids in southeast Asia, \citep{phelps2015invisible}) to overharvesting threats also merits further study. Because price and catch flexibilities ($f, \beta$) can change, a species currently having $f > \beta$ does not necessarily face extinction, absent management, but likely does face economic conditions promoting further depletion that could be severe.

Our analyses assume that species are harvested in an open-access system. Thus, a species we identify as susceptible to extinction from overharvesting would not necessarily be driven extinct if effective management or property rights that promote stewardship were implemented. Indeed, there are property rights in many harvests, including some poaching \citep{harris2013application} and many fisheries \citep{worm2009rebuilding}. Harvests of whales \citep{branch2013opportunistic}, California abalones \citep{hobday2000status}, and bluefin tunas are currently managed; Atlantic bluefin tuna seems to be recovering \citep{pons2016effects}. Trade in sturgeons and many mammals is regulated under CITES \citep{cites2016}. It is nonetheless important to understand how open-access harvesting incentives could lead to extinction, as access restrictions are rarely perfectly enforced; high-value species are especially vulnerable to illegal harvesting \citep{graham2011biodiversity}. 

\matmethods{The \textit{SI Materials and Methods} contains: (i) a glossary of terms, in light of the broad range of topics discussed; (ii) brief discussions of the effects of economies of scale, technological change, rarity effects, supply-independent price trends, perfect substitutes, and non-constant price flexibility ($f$) and catch flexibility ($\beta$) on overharvesting threats (Figs. S2, S3); (iii) descriptions of the range and abundance data shown in Fig. \ref{fig:4}; (iv) discussions of observed range-abundance relationships, in relation to taxonomy and harvesting (Tables S2, S3, Figs. S4, S5); and (v) a review of returns to scale estimates in fisheries (Table S1). Dataset S1 contains all data shown in Figs. \ref{fig:2}, 4, S4, and S5, and their sources.}

\showmatmethods 

\acknow{We thank P. Abrams, C. Lehman, P. Venturelli, D. Williams, and two reviewers for helpful comments; and F. Courchamp, M. Pons, R. Hilborn, T. Branch, A. Scheld, and C. Anderson for assistance with price data. We acknowledge funding from University of Minnesota DDF and NSERC PGS-D fellowships (M.G.B.), the Waitt Foundation, the Ocean Conservancy, and a DoD NDSEG Fellowship (A.F.-H.).}

\showacknow 


\nocite{*}
\bibliography{Burgess-refs.bib}



\section*{Supplementary material}

**Supporting Information for:**
**Range contraction enables harvesting to extinction**
Matthew G. Burgess, Christopher Costello, Alexa Fredston-Hermann, Malin L. Pinsky, Steven D. Gaines, David Tilman, Stephen Polasky

**Contents:**
**SI Materials and Methods**
*Glossary of Key Terms*
*Expanded Theory*
*Economies of scale, technological progress, the rarity effect, and supply-independent price trends*
*Perfect substitutes*
*Non-constant price and catch flexibility (f, β)*
*Data Description*
*Effects of harvesting and taxonomy on the range-abundance relationship*
**Figures S1-S5**
**Tables S1-S3**

**SI Materials and Methods**
*Glossary of Key Terms*
*Hyperstable* catch rates occur when catch-per-unit-effort (CPUE) decreases proportionally slower than abundance as abundance declines, i.e. when catch flexibility ($\beta$) is smaller than 1 (21).

*Rarity value* is consumer value placed on rare species, often because consumption or possession of such species confers status (3).

*Intertemporal discounting* occurs when people consider future consumption or income to be less valuable than equal consumption or income in the present. Economic models often assume that discounting occurs at a constant compounding annual rate (e.g. 3% per year).

*Economies of scale* occur when increasing production inputs results in proportionally larger increases in output. In our context, economies of scale imply that a one-percent increase in harvest effort ($E$)—all else equal—results in a greater-than-one-percent increase in harvest ($Y$) ($\frac{\partial Y}{\partial E}\left[\frac{E}{Y}\right] > 1$).

*Market substitutes*, relative to a focal good, are goods that meet a similar need for consumers, and therefore are goods that consumers would be willing to consume in place of consuming the focal good if prices became more favorable. An increase in the supply of the substitute good causes a reduction in the price of the focal good and vice versa. A *perfect substitute* is indistinguishable from the focal good for consumers, and therefore tends to have the same price (58).

1

*Density-dependent growth* occurs when a biological population's per-capita growth rate changes depending on the population's size. Negative density-dependent growth means that the per-capita growth rate decreases with increasing population size, due to, for instance, intra-species competition.

## *Expanded Theory*
*Economies of scale, technological progress, the rarity effect, and supply-independent price trends*
In the theory presented in the main text, we do not explicitly consider factors other than price and catch flexibility (*f*, *β*) and range changes in driving profitable harvesting to extinction, but we mention several potential factors—economies of scale, technological progress, supply-independent rarity effects, and supply-independent (and positive, as abundance declines) price trends. We claim that economies of scale would have a small mitigating effect on these threats—meaning that conditions on *f* and *β* for profitable harvesting to extinction would be more restrictive—when *f* and *β* were both small (illustrated in Figure S3); and that the others would have an exacerbatory effect (meaning the opposite). We prove these claims below.

Recall the basic general equations of our model, (1)-(3), and extinction condition (4), from the main text:

$$Y = Y(N, E), Y(0, E) = Y(N, 0) = 0 \quad (1)$$

$$\dot{N} = Ng(N) - Y \quad (2)$$

$$\dot{E} > (<) \; 0 \text{ if } p > (<) \; c, \; \dot{E} = 0 \text{ if } p = c \quad (3)$$

$$\lim_{N \to 0} \left\{ \frac{p}{c} \Big|_{Ng(N)=Y} \right\} > 1 \quad (4).$$

Consider expanded price and harvest functions:

$$p = \rho(N, t) Y^{-f} \quad (S1)$$

$$Y = q(t) N^\beta E^\alpha \quad (S2).$$

Now, $\rho$ is a function of abundance and time and $q$ is a function of time, rather than being constant; and harvest (*Y*) is not necessarily linear in *E* (i.e. constant returns to scale, $\frac{\partial Y}{\partial E}\left[\frac{E}{Y}\right] = 1$), and returns to scale are instead given by exponent $\alpha$ (also assumed to be constant, for simplicity) ($\frac{\partial Y}{\partial E}\left[\frac{E}{Y}\right] = \alpha$).

Expanding on extinction condition (4):

$$\lim_{N \to 0} \left\{ \frac{p}{c} \Big|_{Ng(N)=Y} \right\} = r^{\left(1-\frac{1}{\alpha}-f\right)} [q(t)]^{\frac{1}{\alpha}} \lim_{N \to 0} \left\{ \rho(N, t) N^{\left(1+\frac{\beta-1}{\alpha}-f\right)} \right\} > 1 \quad (S3).$$

*Definitions*: I. An economy of scale occurs if $\alpha > 1$. II. If technological progress occurs, $\dot{q} > 0$. III. A supply-independent rarity effect occurs if $\frac{\partial \rho}{\partial N} < 0$. IV. A supply-independent and positive price trend if $\frac{\partial \rho}{\partial t} > 0$.

With these definitions, our claims, formally, are the following:

*Proposition I*: *Suppose extinction condition (10) holds, i.e. $f > \beta$, or $f = \beta$ and $\rho q r^{-f} > 1$, where $\rho = \rho(N[0], 0)$, $q = q(0)$, at the present time ($t = 0$) and abundance (since they are no longer constant as assumed in the main text); the present abundance ($N[0]$) is assumed to be positive, and there are constant returns to scale ($\alpha = 1$). Then, the general extinction condition (S3) holds if any combination of technological progress, the rarity effect, and supply independent and positive price trends (each as defined above) occurs.*

*Proof:*
With constant returns to scale ($\alpha = 1$), condition (S3) becomes:
$$r^{-f} q(t) \lim_{N \to 0} \{\rho(N,t) N^{(\beta-f)}\} > 1 \quad (S4).$$
Our claim is that this condition (S3) holds if (but not only if) $f > \beta$, or $f = \beta$ and $\rho(N[0], 0) q(0) r^{-f} > 1$. Noticing that $\lim_{N \to 0} \rho(N,t) > 0$ because prices are positive and $\frac{\partial \rho}{\partial N} \leq 0$ ($\frac{\partial \rho}{\partial N} < 0$ if there is a rarity effect; $\frac{\partial \rho}{\partial N} = 0$ otherwise), the left-hand side (LHS) of (S4) approaches $\infty$ as $N \to 0$ if $f > \beta$, satisfying the condition in that case. If $f = \beta$ and $\rho(N[0], 0) q(0) r^{-f} > 1$, we can be sure that condition (S4) is satisfied if we can show that:
$$q(t) \lim_{N \to 0} \{\rho(N,t)\} \geq q(0) \rho(N[0], 0) \quad (S5).$$
Because technology is assumed to be either unchanging or progressing ($\dot{q} \geq 0$), we know that $q(t) \geq q(0)$ for all future times ($t > 0$). Because current abundance ($N[0]$) is assumed to be positive, $\frac{\partial \rho}{\partial N} \leq 0$ (possible rarity effect) and $\frac{\partial \rho}{\partial t} \geq 0$ (possible supply-independent and positive price trend), we also know that $\lim_{N \to 0} \{\rho(N,t)\} \geq \rho(N[0], 0)$. Thus, condition (S5)—and by extension also condition (S4)—is satisfied, completing the proof.

*Proposition II*: *For any $f$ and $\alpha$, there is a critical value of $\beta$, denoted $\beta^*$, such that if $\beta < \beta^*$ the general extinction condition (S3) is satisfied with any combination of positive and finite $q(t), r, \lim_{N \to 0} \{\rho(N,t)\}$; and (i) $\beta^* = f$ when $f = 1$; and (ii) $\beta^* < f$ when $\alpha > 1$, and $f < 1$.*

3

*Proof:*
With positive and finite $q(t), r, \lim_{N\to 0}\{\rho(N,t)\}$, the necessary condition for extinction condition (S3) to hold is $\left(1+\frac{\beta-1}{\alpha}-f\right)\leq 0$. Moreover, if $\left(1+\frac{\beta-1}{\alpha}-f\right)<0$, the limit approaches ∞ as $N\to 0$, meaning that,

$$\left(1+\frac{\beta-1}{\alpha}-f\right)<0 \quad (S6),$$

is a sufficient condition for (S3). $\left(1+\frac{\beta-1}{\alpha}-f\right)<0$ if $\beta<\beta^*$, given by:

$$\beta^* = \alpha(f-1)+1 \quad (S7).$$

If $f=1$, $\beta^* = 1 = f$, with any $\alpha$, proving part (i) of *Proposition II*. Differentiating $\beta^*$ with respect to $\alpha$,

$$\frac{d\beta^*}{d\alpha} = f-1 \quad (S8),$$

which, given that $\beta^* = f$ when $f=1$ or $\alpha = 1$, implies that, when $f<1$, $\beta^* < f$ when $\alpha > 1$ (economies of scale) and $\beta^* > f$ when $\alpha < 1$ (Figure S3A) (part [ii] of *Proposition II*).

*Proposition III: For any β and α, there is a critical value of f denoted $f^*$, such that if $f > f^*$ the general extinction condition (S3) is satisfied with any combination of positive and finite $q(t), r, \lim_{N\to 0}\{\rho(N,t)\}$; and (i) $f^* = \beta$ when $\beta = 1$; and (ii) $f^* > \beta$ when $\alpha > 1$, and $\beta < 1$.*

*Proof:*
The proof of *Proposition III* mirrors that of *Proposition II*. Sufficient condition (S6) (for satisfaction of extinction condition [S3]) is satisfied if $f > f^*$, with $f^*$ given by:

$$f^* = 1+\frac{\beta-1}{\alpha} \quad (S9).$$

If $\beta = 1$, $f^* = 1 = \beta$, with any $\alpha$, proving part (i). Differentiating $f^*$ with respect to $\alpha$,

$$\frac{df^*}{d\alpha} = \frac{1-\beta}{\alpha^2} \quad (S10),$$

which, given that $f^* = \beta$ when $\beta = 1$ or $\alpha = 1$, implies that, when $\beta > 1$, $f^* > \beta$ when $\alpha > 1$ (economies of scale) and $f^* < \beta$ when $\alpha < 1$ (Figure S3B) (part [ii]).

Propositions II and III together imply that economies of scale ($\alpha > 1$) have a mitigating effect on the threat of profitable harvesting to extinction when $f, \beta < 1$ because extinction now requires combinations of larger $f$ (i.e. greater price-flexibility) and smaller $\beta$ (lesser catch flexibility). It is worth briefly discussing the intuition of these results concerning economies of scale, which is not immediately obvious. (In contrast, the reasons technological improvement [which makes harvesting cheaper, all else equal], positive price trends, and supply-independent rarity effects should exacerbate the threats should be fairly obvious.)

Positive returns to scale mean—in our framework—that harvest rate ($Y$) is more than linearly sensitive to changes in harvest effort ($E$) (and by extension the costs of harvest, because effort is in monetary units in our model), i.e. $\frac{\partial Y}{\partial E}\left(\frac{E}{Y}\right) = \alpha > 1$. Conversely, this means that requisite harvest effort/total cost ($E$) is less than linearly sensitive to changes in desired harvest rate, i.e. $\frac{\partial E}{\partial Y}\left(\frac{Y}{E}\right) = \frac{1}{\alpha} < 1$. In the extreme version of economies of scale, where $\alpha = \infty$, all harvest rates require the same effort/total cost, i.e. $\frac{\partial E}{\partial Y}\left(\frac{Y}{E}\right) = \frac{1}{\infty} = 0$.

The extinction criterion (conditions [4] and [S3]) is that there has to be upward pressure on effort (i.e. price [$p$] > average costs [$c$]) as abundance approaches 0 ($N \to 0$), at the harvest rates ($Y$) which exactly balance population growth ($Ng[N]$) (i.e. where $Y = Ng[N]$). If abundance ($N$) is small (such that $g(N) \approx r$) (and assuming, for simplicity, that $q$ and $\rho$ are constant), the harvest rate ($Y$), effort/total cost ($E$), average cost ($c$), and price ($p$), where harvest rates exactly balance population growth ($Y = Ng[N]$), are given by:

$$Y_{|Y=Ng(N)} = Ng(N) \approx rN \quad \text{(S11a)},$$

$$E_{|Y=Ng(N)} = q^{-\frac{1}{\alpha}} Y^{\frac{1}{\alpha}} N^{-\frac{\beta}{\alpha}} \approx q^{-\frac{1}{\alpha}} r^{\frac{1}{\alpha}} N^{\frac{1-\beta}{\alpha}} \quad \text{(S11b)},$$

$$c_{|Y=Ng(N)} = q^{-\frac{1}{\alpha}} Y^{\frac{1}{\alpha}-1} N^{-\frac{\beta}{\alpha}} \approx q^{-\frac{1}{\alpha}} r^{\frac{1}{\alpha}-1} N^{\frac{1-\beta}{\alpha}-1} \quad \text{(S11c)},$$

$$p_{|Y=Ng(N)} = \rho Y^{-f} \approx \rho r^{-f} N^{-f} \quad \text{(S11d)}.$$

Under these conditions, if abundance declines by one percent, then the harvest rate ($Y$) exactly balancing population growth ($Ng[N]$) decreases by one percent (from eq. [S11a]); the effort/total cost ($E$) required to achieve this harvest rate changes by ($\beta - 1$)/$\alpha$ percent (from eq. [S11b]), decreasing if $\beta < 1$ and increasing if $\beta > 1$. The intuition here is that, with constant returns to scale ($\alpha = 1$), a one-percent decline in abundance would result in a $\beta$-percent increase in the effort needed to achieve the same harvest rate ($Y$), but since the growth-balancing harvest declines by one percent with a one-percent decline in abundance (assuming abundance is small such that $g(N) \approx r$), the net change in effort ($E$) required is ($\beta - 1$) percent; it is ($\beta - 1$)/$\alpha$ percent with non-constant returns to scale ($\alpha$), where increasing returns to scale ($\alpha > 1$) diminishes the effects of abundance and yield changes on required effort. However, notice that when $\beta = 1$ the abundance and yield effects on the required effort cancel one another out and returns to scale become irrelevant. When $\beta < 1$, the yield effect (reducing the cost of growth-balancing harvest as abundance declines) is larger than the abundance effect (increasing the cost of growth-balancing harvest as abundance declines). By diminishing this source of cost-insensitivity to abundance declines, economies of scale ($\alpha > 1$) have a mitigating effect on threats of profitable harvesting to extinction. Conversely decreasing returns to scale ($\alpha < 1$) would exacerbate the cost insensitivity, and therefore the extinction threats.

Why does the effect of economies of scale on extinction threats also diminish when $f = 1$? When abundance is small such that $g(N) \approx r$, a one-percent decrease in abundance results in a $1 + ([\beta - 1]/\alpha)$-percent change in average costs ($c = E/Y$) of a growth-balancing harvest rate (from eq. [S11c])—coming from a one-percent increase in average costs resulting from a one-percent reduction in the growth-balancing harvest rate (eq. [S11a]), and a $(\beta - 1)/\alpha$-percent change in effort/total costs ($E$) (eq. [S11b])—and an $f$-percent increase in the price (at the growth-balancing harvest rate) via a one-percent decrease in $Y$ (from eq. [S11d]). If $f = 1$, then the direct effects of reducing abundance on costs and prices cancel out, and the net effect on the price/cost ratio is $(\beta - 1)/\alpha$ percent; whether or not this net effect is positive (i.e. price/cost of the growth-balancing harvest rate increases as abundance decreases, which causes extinction [Fig. 1]) is then determined solely by $\beta$, specifically by whether $\beta < 1$. In contrast, if $f, \beta < 1$, the net direct effect of a one-percent reduction in abundance on the price/cost ratio is negative $(f - 1)$, and an extinction threat depends on this effect being more than balanced out by the $(1 - \beta)/\alpha$-percent reduction in total costs; economies of scale ($\alpha > 1$) reduce extinction threats by diminishing this effect.

Given the nuanced and potentially important effect of economies of scale on profit-driven extinction threats, it is worth briefly considering realistic values. In fisheries, where returns to scale have been extensively empirically studied, estimates have tended to be positive ($\alpha > 1$). Across 15 fisheries for which estimates of either $\alpha$ or an equivalent measure of returns to scale (denoted RTS: a composite measure used in studies of multiple factors of production [e.g., see ref. 24]) were available, returns to scale were on average 1.4 (s.d. = 0.50) ($1.40 = \text{RTS} \cong \alpha$) (Table S1). With $\alpha = 1.4$ (and constant $q$ and $\rho$), the extinction criterion becomes, instead of $f \geq \beta$ (condition [10]),

$$f \geq 0.29 + 0.71\beta \quad (S12).$$

Figure S3 compares condition (S12) (dashed black lines) with condition (10) (solid black lines).

*Perfect substitutes*

The existence of substitute goods impacts the sensitivity of a population's harvest price to its supply because it potentially reduces a consumer's willingness to pay a higher price (when they could instead shift demand to the substitute) (33). When a substitute is imperfect—meaning that it meets a similar need as the focal good for consumers, but consumers do not consider the two goods indistinguishable—price flexibility ($f$) can be estimated without bias if the quantity of the substitute good is taken into account, and indeed most studies estimating price flexibility do this (see refs in Dataset S1). As a simple example, suppose the price flexibility of population $i$'s harvest was estimated from an equation of form (S13):

$$\log(p_{it}) = a - f_i \log(Y_{it}) + \sum_k f_{ik} \log(Y_{kt}) + \log(\rho(X_t)) \quad (S13).$$

Here, $\rho(X)$ is an unspecified function of other variables ($X$) impacting prices (e.g., income, etc.), $Y_{kt}$ is the supply of good $k$ in time period $t$, and $f_{ik}$ is the cross-price flexibility (the elasticity of the price of population $i$ to supply changes in good $k$). Good $k$ is a substitute if $f_{ik} < 0$. Provided all $Y_{kt}$ are

6

independent of $Y_{it}$, and no good, $k$, is a perfect substitute of $i$, the relationship, $\frac{\partial p_i}{\partial Y_i}\left(\frac{Y_i}{p_i}\right) = -f_i$, and by extension extinction criterion (10) would still hold.

However, this would not be the case if the harvest of population $i$ had perfect substitutes. Formally, perfect substitutes are defined by the Law of One Price (LOP), which states that perfect substitutes must either have identical or proportional equilibrium prices (where any proportional differences result from differences in quality or transportation costs that are immune to arbitrage) (33, 58). The simplest test for perfect substitutability between two products, denoted 1 and 2, involves estimating the parameters in the following equation:

$$\log(p_1) = a + b\log(p_2) \quad (S14),$$

and testing the hypothesis that $b = 1$ (perfect substitutability). This equation can also be used to test a hypothesis of no market interaction ($b = 0$). Asche *et al*. (33, 58) review other methods for testing for substitutability (often referred to more broadly as 'market integration' in economic studies, as the LOP depends on both equivalence of products for consumers and ease of arbitrage between markets in a spatial sense) that can allow for non-stationarity in prices and other dynamic complexities, when longer unbroken time-series data are available.

Identifying perfect substitutes is important for two reasons. First, if a good, $k$, is a perfect substitute for harvest of a population, $i$ (i.e. the LOP holds), then treating them as separate products creates bias in estimating $f_i$ (even if $\log(Y_k)$ is included in the regression, for example), as the 'true' $f_i$ will vary with the supply of the perfect substitute good at each time period. Second, and more fundamentally, if harvest of population $i$ is part of a set of populations producing equivalent harvests (denoted $S$), then the price responds to changes in the total supply of harvest of all populations in $S$. In other words, price flexibility, $f_i$, would really describe:

$$f_i = \frac{\partial p_i}{\partial \left(\sum_{k=1}^{S} Y_k\right)} \left(\frac{\sum_{k=1}^{S} Y_k}{p_i}\right) \neq \frac{\partial p_i}{\partial Y_i}\left(\frac{Y_i}{p_i}\right) \quad (S15).$$

Given that $Y_i \to 0$ as $N_i \to 0$, it follows from equation (S15) that $(\delta p_i/\delta N_i) \to 0$ as $N_i \to 0$, eliminating the threat of price-driven extinction, if there are other more common populations or products which are perfect substitutes (i.e. $Y_k > 0$, for at least one $k$). Thus, with respect to price-driven extinction threats, a group of populations whose harvests are perfect substitutes are only as weak as their strongest (i.e. most slowly depleted) member. This result may have important implications for rare fish species that are aquacultured. However the assumption that aquaculture is a perfect substitute for wild harvest remains untested for most species, and evidence so far suggests that perfect substitutability may be rare (59).

*Non-constant price and catch flexibility (f, β)*
Consider the original price ($p$) and cost ($c$) functions [(6), (8)] from the main text, and extinction condition (4), but assuming now that $f$ and $β$ are non-constant—for example, potentially dependent on harvest ($Y$), abundance ($N$) or time ($t$).

7

$$p = \rho Y^{-f(Y,N,t)} \quad (S16)$$

$$c = q^{-1} N^{-\beta(Y,N,t)} \quad (S17).$$

$$\lim_{N \to 0} \left\{ \frac{\rho Y^{-f(Y,N,t)}}{q^{-1} N^{-\beta(Y,N,t)}} \bigg|_{Ng(N)=Y} \right\} = \rho q \lim_{N \to 0} \left\{ r^{-f(rN,N,t)} N^{\beta(rN,N,t)-f(rN,N,t)} \right\} > 1 \quad (S18).$$

For extinction condition (S18) to hold, the equivalent of extinction condition (10) ( $f > \beta$, or $f = \beta$ and $\rho q r^{-f} > 1$ ) must hold at the limit as $N \to 0$ where $Y = Ng(N)$, assuming the limit of $f(rN, N, t)$ as $N \to 0$ is finite. One corollary of this result, for example, is that linear demand (i.e. $p = a - bY$, where $a$ and $b$ are constants, implying $f(Y,N,t) = -\frac{\partial p}{\partial Y}\left(\frac{Y}{p}\right) = \frac{bY}{a-bY}$ ) could only allow extinction with either hyperaggregation or abundance-decoupled CPUE at the limit as $N \to 0$ where $Y = Ng(N)$ (i.e. $\lim_{N \to 0} \left\{ \beta(Y,N,t) \big|_{Ng(N)=Y} \right\} \le 0$ ) because $\lim_{N \to 0} \left\{ f(Y,N,t) \big|_{Ng(N)=Y} \right\} = r \lim_{N \to 0} \left\{ \frac{bN}{a-brN} \right\} = 0$.

*Data Description*

The range and abundance data shown in Figs. 4, S4, and S5 and Tables S2 and S3 come from the following sources: Pinsky *et al.* (42) (U.S. fish and invertebrates), Worm and Tittensor (43) (global tunas and billfish range changes), Ricard *et al.* (35) (global tunas and billfish abundance changes), Pillay *et al.* (44) (terrestrial mammals in the Western Ghats of India), Maisels *et al.* (45) (African forest elephant, *Loxodonta cyclotis*), Rodríguez (46) (North American birds).

We compared estimates from U.S. trawl surveys (42) of the range and abundance changes from the 1970s to the 2000s from 367 well-sampled fish and invertebrate populations, each having abundance changes between -100% and 100% (Fig. 4A) (we restricted our sample in this way to remove outliers with very large abundance increases and to give declining and increasing populations equal weight). We calculated range as the number of 0.5°x0.5° grid cells occupied by a species in a given decade. To help correct for false absences, we excluded cells in which a survey did not have a high (≥99%) chance of detecting a species if it was present. We calculated abundance as the stratified mean catch per trawl tow in a decade.

We compared estimates of range changes from the 1960s to the 1990s of 22 tuna and billfish populations (43) to estimates from the RAM Legacy Stock Assessment Database (35) of the abundance change in each population (in units of total biomass) over the same period (Fig. 4A,B, Fig. S4A). Multiple populations of the same species in the same ocean were aggregated to be compatible with Worm & Tittensor's (43) ocean-wide range change estimates. Estimated range changes in SBT were averaged across all 3 oceans (Atlantic, Pacific, Indian).

We compared estimates of range and abundance from 1978-79 to 2008-2009 of 18 terrestrial mammal populations from the Western Ghats of India (44) (Fig. 4A,C, S4B). We used Pillay *et al.*'s (44) measure of occupancy as a proxy for range area, and detectability as a proxy for abundance, as they intended (see their Table 2). We used estimates of percent change in range and abundance reported in

8

Maisels *et al.* (45) for the African forest elephant (2002-2011) (see their Abstract), and reported in Table 2 of Rodríguez (46) for 27 bird populations from North America (1966-1993). Body mass estimates for terrestrial mammals came from Pillay *et al.* (44) for the 18 populations in their analysis, and from University of Michigan Animal Diversity Web (http://animaldiversity.org/accounts/Loxodonta_cyclotis/) for the African forest elephant. For birds, they came from the Cornell Ornithology Lab (https://www.allaboutbirds.org/).

The literature estimates of $\beta$ and $f$ are listed in full in Dataset S1. Estimates of returns to scale, and related references, are listed in Table S1.

*Effects of harvesting and taxonomy on the range-abundance relationship*
In our empirical analysis in the main text of range-abundance relationships (Figs. 2, 4), we only included populations from the above-described samples that are harvested ($n = 160$), despite our data sources also including many species that are not harvested ($n = 275$). For the purposes of this analysis, we defined a marine population as 'harvested' if it had catches reported to the Food and Agriculture Organization of the United Nations (FAO) (http://www.fao.org/fishery/statistics/en)—with this definition identifying 142/389 populations as harvested and 247/389 as not harvested. We note that some populations, which are caught primarily in small-scale recreational, subsistence, or Indigenous traditional harvests, are incorrectly designated not harvested by this method, but there is still a commercial vs. non-commercial harvesting distinction in these cases. We classified the terrestrial mammals and birds as 'harvested' or 'not harvested' according to whether or not their IUCN Red List assessment (1) mentions harvesting as a threat. With this classification, only one of 27 birds is harvested (northern bobwhite [*Colinus virginianus*]) and 17 of the 19 mammals are harvested (dhole [*Cuon alpinus*], golden jackal [*Canis aureus*] are the two populations not harvested).

We briefly examine the full dataset here, to ask: (i) whether harvesting is a significant determinant of the range-abundance relationship, and (ii) whether the broad-scale difference we seem to see in the smaller sample, between the strengths of range-abundance relationships of marine and terrestrial populations, also hold up in the larger sample. In the marine species, both harvested and non-harvested populations had similar taxonomic coverage (the tunas and billfish in our sample—all of which are harvested—were a noteworthy exception).

We tested for an effect of harvesting (our binary definition) on the range-abundance relationship using ordinary least squares (OLS); we regressed the percent change in range area (%$\Delta A$) on combinations of the percent change in abundance (%$\Delta N$), harvested status (a binary variable, as described above), taxonomic group (see Table S3 for marine taxonomic group definitions; terrestrial species were classified simply as either mammals or birds), and their interactions. In all models, we found harvested status, and its interaction with percent change in abundance (%$\Delta N$) (which measures the effect of harvested status on the relationship between range change and abundance change), to be non-significant explanatory variables ($p > 0.05$) (Table S2, Fig. S4). This may suggest that ecological factors (such as aggregation behavior and habitat patchiness) or other human-caused factors (such as habitat loss and climate change) are often the primary drivers of range-abundance relationship, but this

9

hypothesis merits further scrutiny. In Table S3, we compare the fractions of various marine taxa having percent range changes greater than half the their percent abundance changes (%Δ$A$/%Δ$N$ > 0.5). Pelagic finfish (including tunas and billfish) and small invertebrates most commonly had %Δ$A$/%Δ$N$ > 0.5. Differences between demersal and coastal finfish, and tunas and billfish and small molluscs were statistically significant ($p < 0.05$) according to Fisher's exact test.

The significant contrast in the strength of range-abundance relationships of marine vs. terrestrial populations—with stronger relationships among terrestrial populations (Fig. 4A)—persists in the larger sample including both harvested and non-harvested populations (Fig. S5, Dataset S1). Indeed, this contrast persists and is similar when comparing either marine fish or marine invertebrates, to either terrestrial mammals or terrestrial birds (Fig. S5).

**Table S1.** Estimates of returns to scale ($\alpha$, RTS) in fisheries, with references.

| Fishery | Parameter | Value | Reference(s) |
|---|---|---|---|
| NSW Ocean prawn trawl fishery | RTS | 2.6 | Greenville J, Hartmann J, Macauley TG (2006) Technical efficiency in input-controlled fisheries: The NSW ocean prawn trawl fishery. *Mar Resour Econ* 21: 159-179. |
| Alaska pollock (pre-American Fisheries Act) | RTS | 1.95 | Torres MO, Felthoven RG (2014) Productivity growth and product choice in catch share fisheries. *Mar Pol* 50: 280-289. |
| Lofoten (Norway) saithe | RTS | 1.94 | Hannesson R (1983) Bioeconomic production function in fisheries: theoretical and empirical analysis. *Can J Fish Aquat Sci* 40: 968-982. |
| Hawaii longline | RTS | 1.87 | Sharma KR, Leung P (1999) Technical efficiency of the longline fishery in Hawaii: An application of a stochastic production frontier. *Mar Resour Econ* 13: 259-274. |
| Mid-Atlantic sea scallop | RTS | 1.73 | Kirkley JE, Squires D, Strand IE (1995) Assessing technical efficiency in commercial fisheries: The mid-Atlantic sea scallop fishery. *Am J Ag Econ* 77: 686–697. |
| Mid-Atlantic ocean quahog | RTS | 1.51 | Weninger Q (1998) Assessing efficiency gains from individual transferable quotas: An application to the mid-Atlantic surf clam and ocean quahog fishery. *Am J Ag Econ* 80: 750-764. |
| Iranian Persian Gulf fishery | RTS | 1.42 | Esmaeili A (2006) Technical efficiency analysis for the Iranian fishery in the Persian Gulf. *ICES J Mar Sci* 63: 1759-1764. |
| North Sea herring (24, 73) | $\alpha$, RTS | 1.40[*] | Bjørndal T, Conrad JM (1987) The dynamics of an open access fishery. *Can J Econ* 20: 74-85. Bjorndal T (1989) Production in a schooling fishery: The case of the North Sea herring fishery. *Land Econ* 65: 49-56. |

| Alaska pollock (post-American Fisheries Act) | RTS | 1.07 | Torres MO, Felthoven RG (2014) Productivity growth and product choice in catch share fisheries. *Mar Pol* 50: 280-289. |
|---|---|---|---|
| Mid-Atlantic surf clam | RTS | 1.04 | Weninger Q (1998) Assessing efficiency gains from individual transferable quotas: An application to the mid-Atlantic surf clam and ocean quahog fishery. *Am J Ag Econ* 80: 750-764. |
| Lofoten (Norway) cod | RTS | 1.06 | Hannesson R (1983) Bioeconomic production function in fisheries: theoretical and empirical analysis. *Can J Fish Aquat Sci* 40: 968-982. |
| Pacific halibut | RTS | 1 | Comitini S, Huang DS (1967) A study of production and factor shares in the halibut fishing industry. *J Poli Econ* 75: 366-372. |
| Australian Southeast trawl fishery | RTS | 0.99 | Kompas T, Che TN (2005) Efficiency gains and cost reductions from individual transferable quotas: A stochastic cost frontier for the Australian south east fishery. *J Prod Anal* 23: 285-307. |
| Solomon Islands pole and line fishery | RTS | 0.96 | Campbell HF, Hand AJ (1998) Joint ventures and technology transfer: the Solomon Islands pole-and-line fishery. *J Devel Econ* 57: 421-442. |
| Swedish trawl fishery for Norway lobster | RTS | 0.92 | Eggert H (2000) Technical efficiency in the Swedish trawl fishery for Norway lobster. *IIFET 2000 Proceedings*. Available from http://oregonstate.edu/dept/IIFET/2000/papers/eggert.pdf 448 (Accessed April 25, 2016). |
| NE Atlantic minke whale | $\alpha$ | 0.865 | Amundsen ES, Bjorndal T, Conrad JM (1995) Open access harvesting of the northeast Atlantic minke whale. *Environ Res Econ* 6: 167-185. |
| **Mean (S.D.)** | | **1.40 (0.50)** | |

[*]Mean of two estimates.

**Table S2.** Results of OLS regressions of range change (%Δ$A$) on combinations of: abundance change (%Δ$N$), harvested status (for marine species: 'Y' if species has an FAO capture record, 'N' otherwise; for terrestrial species: 'Y' if species' IUCN Red List assessment lists harvesting as a threat, 'N' = otherwise), taxonomic group (defined as in Table S3 for marine species, 'mammal' or 'bird' for terrestrial species), and interactions. Marine and terrestrial species are modeled separately, and two outliers are excluded in the terrestrial analysis (lark sparrow [*Chondestes grammacus*] and Brewer's blackbird [*Euphagus cyanocephalus*], similarly to Figs. S4 and S5). Each row in the table represents a separate regression model, with its explanatory variables indicated (italicized explanatory variables had significant effects on range change [%Δ$A$] [$p < 0.05$]), and the effect size (on range change [%Δ$A$]) of harvesting and its interaction with abundance change (%Δ$N$) indicated along with the corresponding $p$ value. In all models, the harvesting effect was not found to be significant.

| Marine/Terrestrial | Explanatory Variables | Harvesting Effect [N] ($p$ value) | (Harvesting [N]) x (%Δ$N$) Effect ($p$ value) |
|---|---|---|---|
| Marine | *Abundance change (%Δ$N$)*, Harvested? | -0.36 (0.57) | N/A |
| Marine | *Abundance change (%Δ$N$)*, Harvested?, (%Δ$N$) x Harvested? | -0.34 (0.58) | 0.017 (0.14) |
| Marine | *Abundance change (%Δ$N$)*, *Taxonomic Group*, Harvested? | -0.045 (0.50) | N/A |
| Marine | *Abundance change (%Δ$N$)*, *Taxonomic Group*, Harvested?, (%Δ$N$) x Harvested? | -0.41 (0.53) | -0.016 (0.16) |
| Marine | *Abundance change (%Δ$N$)*, *Taxonomic Group*, (%Δ$N$)xTaxonomic Group, Harvested? | -0.62 (0.34) | N/A |
| Terrestrial | *Abundance change (%Δ$N$)*, Harvested? | -0.85 (0.56) | N/A |
| Terrestrial | *Abundance change (%Δ$N$)*, Harvested?, (%Δ$N$) x Harvested? | -0.73 (0.61) | -0.098 (0.30) |
| Terrestrial | *Abundance change (%Δ$N$)*, *Taxonomic Group*, Harvested? | 4.5 (0.07) | N/A |
| Terrestrial | *Abundance change (%Δ$N$)*, *Taxonomic Group*, Harvested?, (%Δ$N$) x Harvested? | 4.2 (0.088) | -0.051 (0.57) |

| Terrestrial | *Abundance change (%ΔN), Taxonomic Group*,  (%ΔN) x Taxonomic Group, Harvested? | 4.5 (0.087) | N/A |

**Table S3.** Distributions of range-abundance relationships (measured by %Δ$A$/%Δ$N$) among marine populations by taxonomic group (sorted by fraction of populations with %Δ$A$/%Δ$N$ > 0.5). Superscript letters denote pairwise significant differences ($p$ < 0.05) between taxonomic groups in the fraction of populations with %Δ$A$/%Δ$N$ > 0.5, calculated using Fisher's exact test. The "Small pelagic finfish" category includes FAO ISSCAAP (ftp://ftp.fao.org/FI/STAT/DATA/ASFIS_structure.pdf) categories "Shads" and "Herrings, sardines, anchovies"; the "Demersal and coastal finfish" category includes FAO ISSCAAP categories "Cods, hakes, haddocks", "Flounders, halibuts, soles", "Salmons, trouts, smelts", "Miscellaneous coastal fishes", and "Miscellaneous demersal fishes".

| Taxonomic Group | Number of Populations | Fraction with %Δ$A$/%Δ$N$ > 0.5 | Median %Δ$A$/%Δ$N$ (25th pct., 75th pct.) |
|---|---|---|---|
| Molluscs (excl. squids, cuttlefishes, octopuses) | 18 | 0.28[a,b] | 0.11 (0.04, 0.58) |
| Tunas, billfishes | 23 | 0.17[a,b] | 0.08 (0.00, 0.31) |
| Small pelagic finfish | 12 | 0.17[a,b,c] | 0.10 (−0.02, 0.34) |
| Other | 13 | 0.15[a,b,c] | 0.00 (−0.13, 0.14) |
| Other pelagic finfish | 38 | 0.13[a,b,c] | 0.01 (−0.04, 0.20) |
| Shrimps, prawns | 21 | 0.095[a,b,c] | 0.00 (−0.03, 0.10) |
| Demersal and coastal finfish | 230 | 0.052[c] | 0.00 (−0.01, 0.09) |
| Other crustaceans | 23 | 0[b,c] | 0.00 (−0.06, 0.03) |
| Squids, cuttlefishes, octopuses | 11 | 0[a,b,c] | 0.00 (−0.05, 0.14) |

14

**Figures**

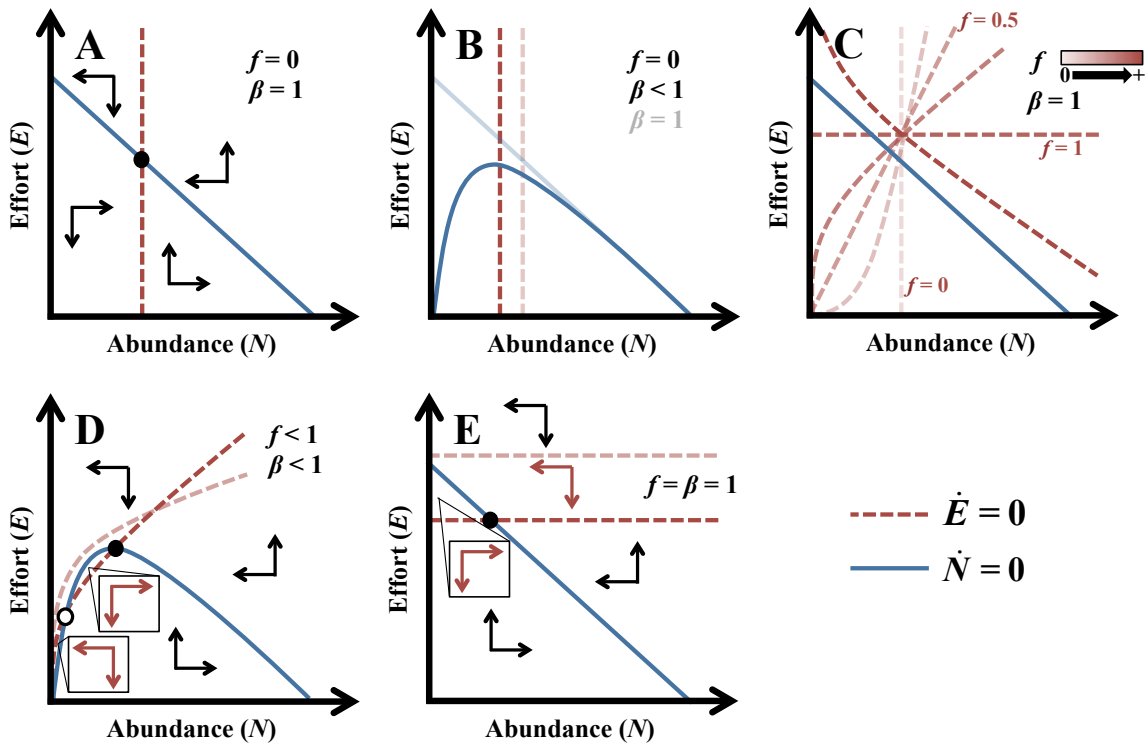

**Figure S1.** Isoclines and directions of change depicted in *N-E* phase plane for different values of $\beta$ and *f*. Panel **A** illustrates the general concept of the isoclines; panels **B** and **C** illustrate the effects on the isoclines of varying values of $\beta$ and *f*, respectively; and panels D and E illustrate scenarios in which extinction is possible, where $f > \beta$ (**D**) or $f = \beta$ (**E**). Open and filled circles in **A**, **D** and **E** denote unstable and potentially stable (cycling) equilibria, respectively. Because both *E* and *N* are bounded and the phase plane is two-dimensional, the dynamics of *N* and *E* must eventually reach the origin (extinction), the filled equilibrium point, or a limit cycle. Extinction is possible when the *E* isocline (red) lies above the *N* isocline at the limit when *N* approaches zero (equivalent to condition [4] in the main text).

15

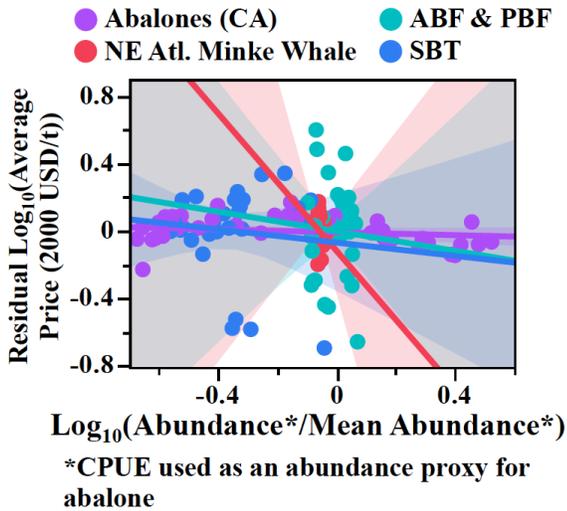

**Figure S2.** Historical relationships between price ($p$) and abundance ($N$), controlling for production ($Y$), for four of the populations analyzed in Fig. 3 (caviar not shown here due to lack of abundance information). None of these relationships were significant, suggesting price flexibility ($f$) may capture rarity effects for these products.

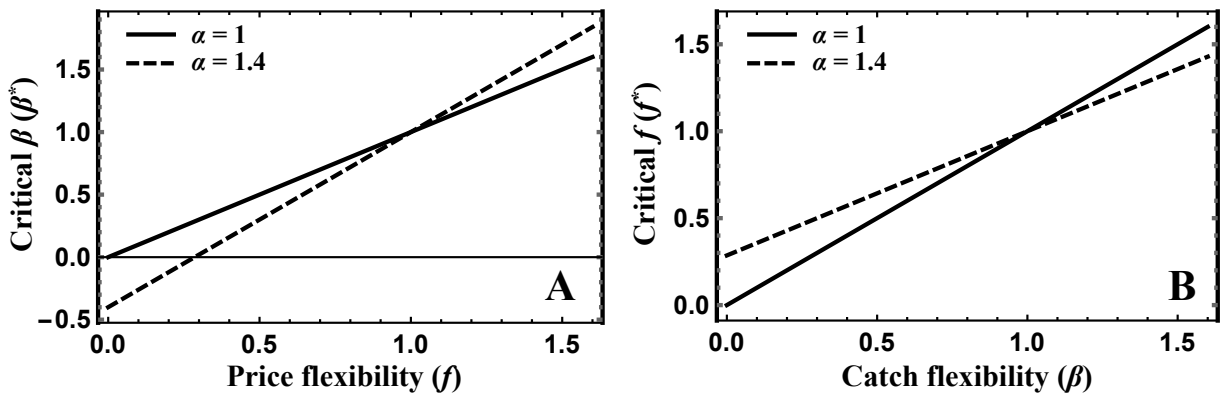

**Figure S3.** The impact of economies of scale ($\alpha > 1$) on the critical values of (**A**) $\beta$ and (**B**) $f$ needed for an extinction, illustrated with $\alpha = 1.4$ (dashed, see Table S1) and $\alpha = 1$ (solid, constant returns to scale).

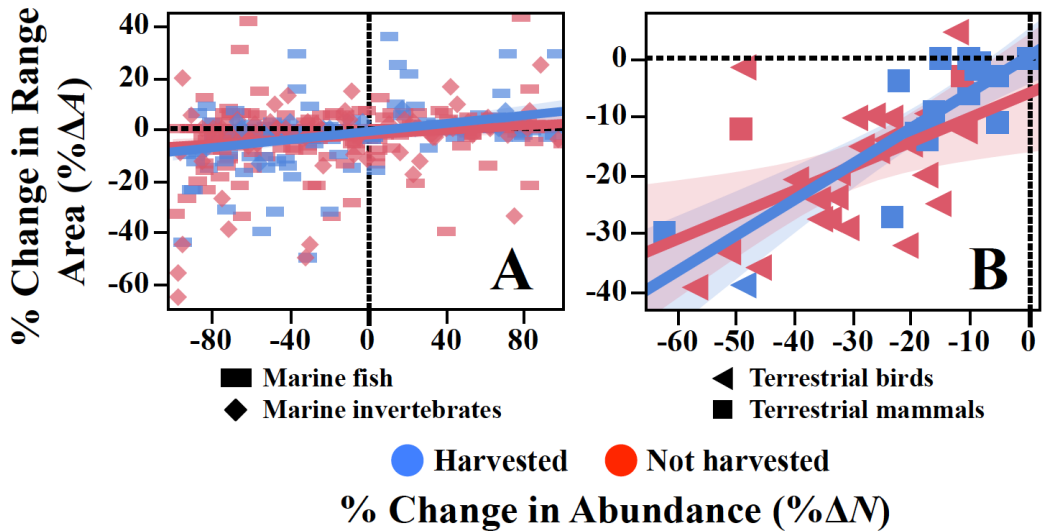

**Figure S4.** Range-abundance relationships of (**A**) marine fish and invertebrates, and (**B**) terrestrial mammals and birds, comparing harvested (blue) and non-harvested (red) populations. Lines indicate linear OLS fits within each group (harvested vs. not harvested), with 95% confidence intervals shaded. Two outliers among the birds were excluded in these fits, as indicated in Dataset S1.

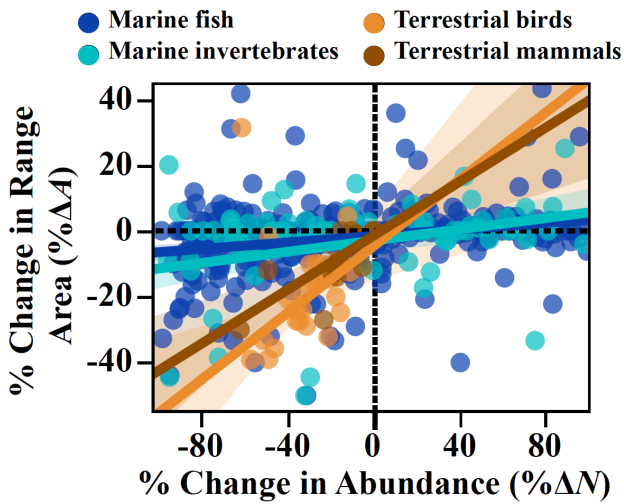

**Figure S5.** Range-abundance relationships of all populations—both harvested and not harvested—for which data were available. Lines indicate linear OLS fits within each large taxonomic group, with 95% confidence intervals shaded, as indicated. Two outliers among the birds were excluded in these fits, as indicated in Dataset S1.



\end{document}